\documentclass[11pt]{article}

\usepackage[margin=1in]{geometry}

\usepackage{graphicx}
\usepackage[utf8]{inputenc} %
\usepackage[T1]{fontenc}    %
\usepackage{hyperref}       %
\usepackage{url}            %
\usepackage{booktabs}       %
\usepackage{amsfonts}       %
\usepackage{nicefrac}       %
\usepackage{microtype}      %
\usepackage{verbatim}

\usepackage{multirow}
\usepackage{multicol}
\usepackage{enumitem}
\usepackage{bm}%
\usepackage{amsmath}%
\usepackage{amssymb}%
\usepackage{amsthm}
\usepackage{colonequals}
\usepackage{psfrag}
\usepackage{mathrsfs}
\usepackage{booktabs,siunitx}
\usepackage{algorithm}
\usepackage{algpseudocode}

\usepackage[usenames,dvipsnames]{xcolor}

\hypersetup{
    colorlinks,
    linkcolor={blue},%
    citecolor={green!60!black},
    urlcolor={blue}%
}
\usepackage[square,numbers,sort&compress]{natbib}

\newtheorem{thm}{Theorem}
\newtheorem{lem}{Lemma}

\theoremstyle{definition}

\newtheorem{defn}{Definition}

\title{Private Synthetic Data Meets Ensemble Learning}

\author{%
    Haoyuan Sun$^1$, Navid Azizan$^1$, Akash Srivastava$^2$, and Hao Wang$^2$ \\
    \\
    $^1$ {\footnotesize Massachusetts Institute of Technology} \\
    $^2$ {\footnotesize MIT-IBM Watson AI Lab}
}

\date{}

\begin{document}

\maketitle

\renewcommand{\thefootnote}{\roman{footnote}}
\footnotetext[0]{Corresponding author: Haoyuan Sun (\texttt{haoyuans (at) mit (dot) edu}). Working paper. Comments to improve this work are welcome.}
\renewcommand{\thefootnote}{\arabic{footnote}}

\vspace{-.2in}

\begin{abstract}\noindent
When machine learning models are trained on synthetic data and then deployed on real data, there is often a performance drop due to the distribution shift between synthetic and real data. In this paper, we introduce a new ensemble strategy for training downstream models, with the goal of enhancing their performance when used on real data. We generate multiple synthetic datasets by applying a differential privacy (DP) mechanism several times in parallel and then ensemble the downstream models trained on these datasets. While each synthetic dataset might deviate more from the real data distribution, they collectively increase sample diversity. This may enhance the robustness of downstream models against distribution shifts. Our extensive experiments reveal that while ensembling does not enhance downstream performance (compared with training a single model) for models trained on synthetic data generated by marginal-based or workload-based DP mechanisms, our proposed ensemble strategy does improve the performance for models trained using GAN-based DP mechanisms in terms of both accuracy and calibration of downstream models.
\end{abstract}

\section{Introduction}

The success of modern machine learning (ML) techniques hinges on access to vast amounts of high-quality training data. However, data sharing presents significant challenges due to inherent privacy risks. In 2023, 19\% of data breaches were attributed to compromises with business partners, and in the U.S., the average cost of a data breach reached \$9.44 million \cite{james2023databreach}. Additionally, stringent data protection regulations, such as GDPR \cite{gdpr} and CCPA \cite{ccpa}, restrict the external sharing of data or its transfer across geographic boundaries. Therefore, there is both a business and scientific imperative to develop solutions that allow data curators to share data reliably (without worrying about privacy leakage) and efficiently (with streamlined end-to-end implementation). Synthetic data generation offers a promising solution and contemporary generative models can produce content that is very close, or even indistinguishable, in quality to human-made content. On the other hand, generative models might inadvertently memorize and reproduce specific details from the training data \citep{hayes2017logan,chen2020gan,jordon2022synthetic}. Consequently, it is crucial to train these models in compliance with privacy standards.

A burgeoning body of research is focused on developing differentially private (DP) synthetic data generation mechanisms \citep[see, e.g.,][]{wang2023post,mckenna2021winning,tao2021benchmarking,ullman2020pcps,abay2019privacy,ge2020kamino,liu2021iterative,ping2017datasynthesizer,ridgeway2021challenge}. 
They can be broadly classified into three categories: (i) GAN-based mechanisms train generative adversarial networks with DP guarantees \citep{xie2018differentially,beaulieu2019privacy,jordon2019pate,tantipongpipat2019differentially,neunhoeffer2020private,uclaneslDPwgan}; (ii) marginal-based mechanisms select several subsets of features, compute their joint distributions, add noise, and then fit a graphical model or neural network to match the noisy distributions \citep{zhang2017privbayes,mckenna2019graphical,mckenna2021winning}; (iii) workload-based mechanisms select a set of workload queries and generate synthetic data aimed at minimizing the approximation error for the chosen queries \citep{vietri2020new,aydore2021differentially,mckenna2022aim,liu2021iterative,vietri2022private,liu2023generating}. 
Despite the differences, all these mechanisms share a common goal: produce synthetic data that retains key information (such as statistical properties) from the original private dataset while ensuring that adversaries cannot identify individual personal data from the released synthetic data.
A significant application of synthetic data is to allow downstream users to train their ML models, with the expectation that these models will maintain their performance when applied to real data.
\vspace{0.5em}

Existing studies have observed that ML models trained on private synthetic data can underperform when applied to real data \citep{tao2021benchmarking,wang2023post}. This performance gap arises because synthetic data typically follows a different probability distribution than real data. Several factors can cause this distribution shift, including noise introduced to ensure DP guarantees, the limited size of training data, and inherent limitations of generative models or the training paradigm. Additionally, synthetic data may not capture the full diversity of real data because generative models tend to avoid producing data from domains with limited real data to meet privacy constraints. Yet, these outliers can be pivotal for training certain downstream models, such as SVMs. %
Given these challenges, it is crucial to develop a systematic approach for training downstream models with synthetic data while addressing distribution shifts and lack of sample diversity.

Ensemble learning trains multiple predictive models and aggregates their predictions to achieve more accurate and reliable outcomes than individual models can provide. 
At the heart of ensemble methods is to ensure diversity among individual models. %
In our context---training downstream models using synthetic data---we identify an additional source to amplify model diversity: the noise introduced in DP synthetic data generation mechanisms. 
By applying a privacy mechanism multiple times in parallel to real data, we generate synthetic data with diverse distributions. This approach is particularly beneficial for GAN-based privacy mechanisms, which are prone to mode collapse and hard to fine-tune due to privacy constraints.
We demonstrate that injecting different noise in GAN-based mechanisms can significantly improve sample diversity and enable downstream models to be more robust against distribution shift. As a result, these models are more likely to maintain their performance when deployed on real data.

\begin{figure}
    \centering
    \includegraphics[width=0.98\textwidth]{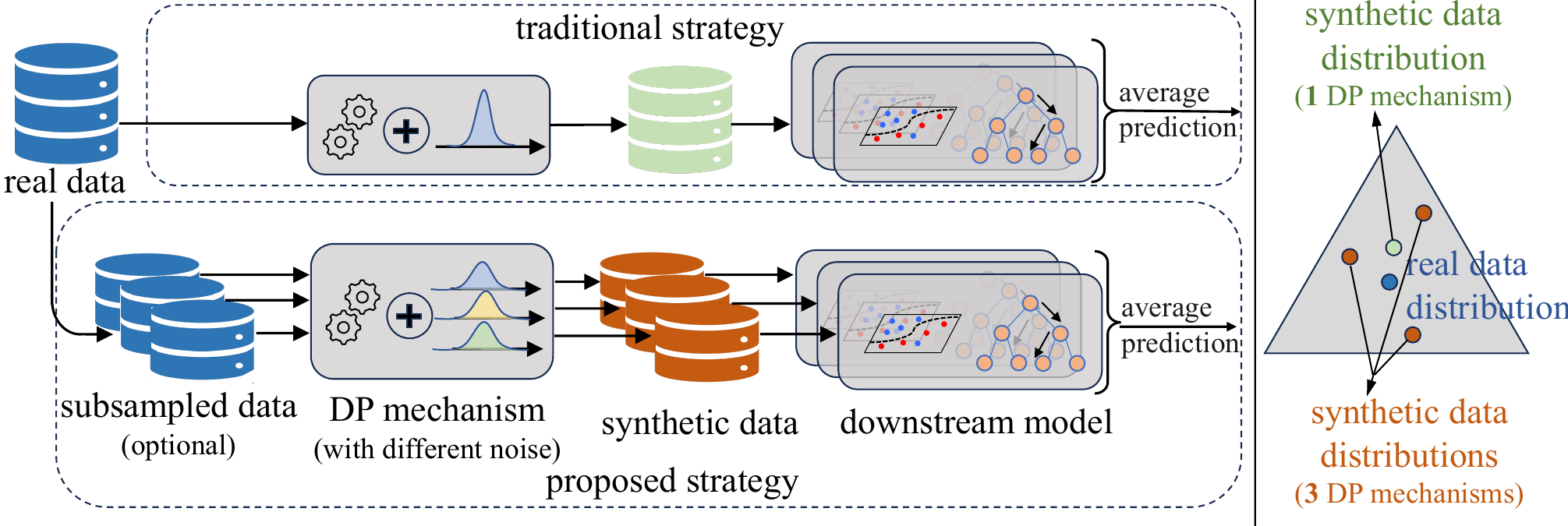}
    \caption{Illustration of our proposed method. 
    \textit{Left:} In our proposed strategy, we fix a predefined privacy budget, apply a DP mechanism multiple times in parallel to generate various synthetic datasets, and ensemble the downstream models trained on each of these synthetic datasets. If real data exhibits sufficient diversity, we also subsample it before applying the DP mechanism, saving privacy budget. 
    \textit{Right:} Interpretation of the proposed strategy. Although each synthetic data distribution might diverge farther from the real data distribution, they increase sample diversity, which in turn benefits downstream models by making them more robust against distribution shifts.}
    \label{fig:pipeline}
    \vspace{2em}
\end{figure}

We introduce a new strategy to ensemble downstream models (see Figure~\ref{fig:pipeline}), aiming at improving their predictive accuracy and calibration when deployed on real data. 
Specifically, we fix a predefined privacy budget, divide it among several DP synthetic data generation mechanisms, and apply each mechanism to produce synthetic data. We then evaluate the performance of downstream models trained on this ensemble dataset against models trained on synthetic data generated using the entire privacy budget. We conduct comprehensive numerical experiments using state-of-the-art DP mechanisms and various benchmark datasets. Our key observations are: 
\begin{itemize}[leftmargin=*]
    \item For marginal-based and workload-based mechanisms, ensemble methods do not improve the performance of downstream models compared to training just one model. This holds true whether the ensemble models are trained on a single synthetic dataset or on different synthetic datasets generated through multiple invocations of a DP mechanism.
    
    \item For GAN-based mechanisms, our proposed ensemble strategy improves both the accuracy and calibration of downstream models. In particular, we find that taking a random sub-sample of the real data before each invocation of GAN-based mechanisms is particularly beneficial to improve downstream model performance.
\end{itemize}

\subsection*{Related Work}

Ensemble learning is a standard strategy in supervised learning, aimed at enhancing model accuracy and reliability \citep{dietterich2000ensemble}. Traditional methods include bagging~\cite{breiman1996bagging}, boosting~\cite{freund1997decision}, and Bayesian model averaging~\cite{hoeting1999bayesian}. For deep learning models, there are also advanced ensemble techniques such as Monte-Carlo (MC) dropout~\cite{gal2016dropout} and deep ensemble~\cite{lakshminarayanan2017simple}. To encourage diversity among individual models, various strategies have been suggested, for instance, deliberately manipulating the randomness during the model training process or training each model using a random subset of the available data and features \cite{breiman2001random}. 
However, when these methods are applied to synthetic data, their effectiveness hinges on the quality of the generative model. A mode collapse in the generative model can drastically diminish the diversity of training samples, subsequently affecting the diversity of the ensemble models. In contrast, we suggest repeatedly applying DP mechanisms (within a set privacy budget) to produce more diverse synthetic data. Our experiments (Figure~\ref{fig:gan}) show that for GAN-based privacy mechanisms, this method can  improve both the accuracy and calibration of models compared to the naive application of ensemble learning techniques.

Our paper is inspired by a remarkable work \cite{van2023synthetic}, which proposed learning multiple generative models to produce diverse synthetic datasets, training individual downstream models on each dataset, and aggregating their predictions. They empirically observed that this simple approach achieves superior downstream performance and uncertainty quantification than training with a single synthetic dataset. 
However, when considering privacy constraints, each generation of synthetic data could potentially expose private information from the original dataset. Given a set privacy budget, producing more private synthetic datasets means allocating a smaller privacy budget to each DP synthetic data generation mechanism, consequently affecting the quality of the produced datasets. 
In our experiments, we find that the benefits of using multiple generative models under a fixed privacy budget differ based on the privacy mechanisms employed. For marginal-based and workload-based mechanisms, the gains in downstream performance are negligible (see Figure~\ref{fig:marginal}). 
This observation aligns with the ``no free lunch'' result in \cite{abe2022deep}, which considers uncertainty quantification of models trained on real data and demonstrates that an ensemble of smaller neural networks often performs similarly to a larger neural network under comparable computational costs.
On the other hand, for GAN-based mechanisms, an ensemble with multiple generative models improves downstream performance (see Figure~\ref{fig:gan}).

\section{Preliminaries: Differential Privacy}
\label{sec::pre}

We first recall the definition of differential privacy (DP) \citep{dwork2014algorithmic}. 
\begin{defn}
A randomized mechanism $\mathcal{M}: \mathcal{X}^n \to \mathcal{R}$ satisfies $(\epsilon, \delta)$-differential privacy, if for any adjacent datasets $\mathcal{D}$ and $\mathcal{D}'$, which only differ in one individual's record, and all possible events from the mechanism $\mathcal{O}\subseteq \mathcal{R}$, we have
\begin{align}
    \Pr(\mathcal{M}(\mathcal{D}) \in \mathcal{O})
    \leq e^{\epsilon} \Pr(\mathcal{M}(\mathcal{D}') \in \mathcal{O}) + \delta.
\end{align}
\end{defn}

Within this definition, the \textit{privacy budget}~$\epsilon$ balances the privacy-utility trade-off (a smaller value of $\epsilon$ offers a stronger privacy guarantee); the $\delta$ term allows for a small probability that the privacy guarantee is violated. Typically, $\delta$ is chosen to be a sufficiently small number ($\delta \ll 1/n$, where $n$ is the size of the dataset). 

DP has many compelling properties. The post-processing property states that applying any (potentially randomized) function to the output from a DP mechanism  does not compromise the privacy guarantee. The basic composition rule allows us to compute the privacy guarantee when we apply multiple DP mechanisms.
\begin{lem}[Post-processing]
\label{lem:post-processing}
    If a randomized mechanism $\mathcal{M}: \mathcal{X}^n \to \mathcal{R}$ is $(\epsilon, \delta)$-differentially private, then for any randomized mapping $\mathcal{F} : \mathcal{R} \to \mathcal{R}$, the mechanism $\mathcal{F} \circ \mathcal{M}$ is also $(\epsilon, \delta)$-differentially private.
\end{lem}
\begin{lem}[Basic Composition Rule]
\label{lem:composition}
    Consider a sequence of randomized mechanism $\mathcal{M}_1, \dots \mathcal{M}_k: \mathcal{X}^n \to \mathcal{R}$ and $\mathcal{M}_i$ are $(\epsilon_i, \delta)$-differentially private, respectively.
    Then, the mapping $\mathcal{M}(x) = (\mathcal{M}_1(x), \dots, \mathcal{M}_k(x))$ is $(\sum_{i=1}^k \epsilon_i , \sum_{i=1}^k \delta_i)$-differentially private.
\end{lem}

Another fundamental property of DP is privacy amplification by subsampling. It suggests that the privacy guarantees of a DP mechanism can be enhanced by applying the mechanism to a random subsample of the original dataset \citep{kasiviswanathan2011can,li2012sampling}. This property can be particularly useful if computational or memory constraints restrict the use of a DP mechanism on the entire dataset. Below, we recall a result regarding Poisson subsampling, and for more on different subsampling schemes, refer to \citep{balle2018privacy}.

\begin{lem}
\label{lem:amplification}
Given a dataset $\mathcal{D} = \{x_1,\cdots,x_n\}$, $\mathsf{PoissonSample}_p$ outputs a subset $\{x_i \mid \sigma_i = 1, i\in[n] \}$ such that $\sigma_i$ is independently drawn from a Bernoulli distribution with mean $p$. If mechanism $\mathcal{M}$ is $(\epsilon, \delta)$-DP, then $\mathcal{M} \circ \mathsf{PoissonSample}$ satisfies $(\epsilon', \delta')$-DP where $\epsilon' = \log(1 + p(e^\epsilon-1))$, $\delta' = p\delta$.
\end{lem}

\section{Different Strategies for Ensembling Downstream Models}
\label{sec::ensemb_strategy}

In this section, we introduce a range of ensemble strategies designed for training downstream models using private synthetic data. Our aim is to ensure that these models remain robust against distribution shifts, thereby maintaining their performance when applied to the original data. We start with revisiting two traditional strategies for training models before delving into our main method ``\textit{DP ensemble}'' and its variation.
\begin{itemize}[leftmargin=*]
    \item \textbf{Without ensemble}: We apply a single $(\epsilon, \delta)$-DP mechanism to generate a synthetic dataset. Subsequently, we train a single predictive model on this synthetic dataset.

    \item \textbf{Model ensemble}: We apply a single $(\epsilon, \delta)$-DP mechanism to generate a synthetic dataset. Subsequently, we train $k$ predictive models on this synthetic dataset and aggregate their outputs. To enhance model diversity, we resample a subset from the synthetic data before training each model. This strategy is depicted in the first pipeline of Figure~\ref{fig:pipeline}.
\end{itemize}
A limitation of these traditional approaches is that their effectiveness hinges on the quality of the synthetic data generated by the privacy mechanism. In practice, GAN-based privacy mechanisms are prone to mode collapse and can be challenging to fine-tune due to privacy constraints.  If synthetic data fails to accurately capture the real data distribution, the efficiency of downstream models may decrease, even with multiple models in an ensemble.

Next, we introduce ``\textit{DP ensemble}'' and its variant (see the second pipeline of Figure~\ref{fig:pipeline} for an illustration). This approach applies a privacy mechanism multiple times to generate a variety of synthetic datasets. Although the distribution of each individual dataset may diverge more from the original data distribution than a dataset generated from a single run of the privacy mechanism, collectively they demonstrate a higher sample diversity. This increased diversity can enhance downstream models, making them more robust against distribution shifts.

\begin{itemize}[leftmargin=*]
    \item \textbf{Simple DP ensemble:} We apply an $(\epsilon/k, \delta/k)$-DP mechanism $k$ times to generate $k$ distinct synthetic datasets. Subsequently, we train a predictive model on each synthetic dataset and aggregate their outputs.
    
    \item \textbf{DP ensemble + subsampling:} We apply $\mathsf{PoissonSample}_{p}$ to subsample from the original real data, apply a $(\log(1 + (\exp(\epsilon/k)-1)/p), \delta/pk)$-DP mechanism to the subsampled data, and repeat this process $k$ times to generate $k$ distinct synthetic datasets. Subsequently, we train a predictive model on each synthetic dataset and aggregate their outputs. 
    
\end{itemize}
Subsampling from the original data before applying the DP mechanism enhances the diversity of synthetic data produced in different runs of the DP mechanism. Additionally, it increases the privacy budget for each run. For example, given $\epsilon = 3$, $k = 3$, and $p = 0.2$, subsampling amplifies the privacy budget for each DP mechanism invocation from $1$ to $2.26$. Our experiments (Figure~\ref{fig:gan}) indicate that subsampling is especially beneficial for GAN-based privacy mechanisms. Specifically, the calibration error of downstream models, when trained with private synthetic data, diminishes when we employ the ``DP ensemble + subsampling'' approach for training ensemble models.

We end this section by establishing a DP guarantee for the aforementioned ensemble strategies.
\begin{thm}
The above four ensemble strategies all satisfy $(\epsilon,\delta)$-DP no matter how many downstream models are trained on the private synthetic data.
\end{thm}

\section{Numerical Experiments}
We conduct comprehensive numerical experiments to evaluate ensembling strategies proposed in Section~\ref{sec::ensemb_strategy}. We provide details about our experimental setup and then present our main observations. We defer additional experimental results to Appendix~\ref{append::exp}.

\vspace{-0.5em}
\paragraph{Dataset.}
We select three tabular datasets from the UCI machine learning repository \cite{misc_uci_repo}: \texttt{Adult}, \texttt{Bank-Marketing}, and \texttt{Online-Shoppers}. All the datasets have both categorical and numerical features and a binary target variable that can be used for a downstream classification task. 
Because marginal-based DP mechanisms only work with discrete data, we convert all numerical features to ordinal values using \texttt{OpenDP}'s \cite{opendp2023} default pre-processing mechanism with a privacy budget $\epsilon = 1$.

\vspace{-0.5em}
\paragraph{DP mechanism.}

We consider five state-of-the-art DP mechanisms to generate synthetic data: \texttt{MST} \cite{mckenna2021winning}, \texttt{MWEM} \cite{hardt2012simple}, \texttt{GEM} \cite{liu2021iterative}, \texttt{DP-CTGAN} \cite{rosenblatt2020differentially} and \texttt{PATE-CTGAN} \cite{rosenblatt2020differentially}. Among them, \texttt{MST} is a marginal-based mechanism; \texttt{MWEM} and \texttt{GEM} are workload-based mechanisms with workload queries being low-order marginals; and \texttt{DP-CTGAN} and \texttt{PATE-CTGAN} are GAN-based mechanisms. All of their implementations, except \texttt{GEM}, are from the OpenDP library \cite{opendp2023}. For \texttt{GEM}, we use the code provided by its authors\footnote{\url{https://github.com/terranceliu/iterative-dp}}. 
We assign a total privacy budget $\epsilon = 3.0$ for running these DP mechanisms. We apply DP ensemble by running the DP mechanism $k = 3$ times, resulting in $3$ synthetic datasets. When we apply DP ensemble + subsampling, we set the subsampling rate as $p = 0.2$ to have a strong privacy amplification effect.

\vspace{-0.5em}
\paragraph{Downstream model.}
We consider four different downstream classifiers: logistic regression, SVM with RBF kernel, random forest, and gradient-boosted decision tree. We train these predictive models using the synthetic data and then test their performance on real data. 
We consider two metrics: accuracy and expected calibration error (ECE). The latter measures the uncertainty of a model by taking the difference between its prediction confidence and its true prediction accuracy.
In practice, ECE is approximated by binning the confidence probabilities into discrete intervals of equal width and assigning each data point to a bin based on the model's prediction confidence:
\begin{align*}
    \mathrm{ECE} = \sum_{b=1}^B \frac{b_n}{N}|\mathrm{acc}(b) - \mathrm{conf}(b)|,
\end{align*}
where $b_n$ is the number of data points falling under bin $b$ and $\mathrm{acc}(b), \mathrm{conf}(b)$ are respectively the accuracy and average confidence on bin $b$.
In order to compute ECE, the model must output a confidence probability for each prediction. 
Then, the ensemble's output is the average of its member classifiers' confidence.
Because an SVM model does not naturally support confidence outputs, we use Platt's scaling~\cite{platt1999probabilistic} to convert the model's margin into a probabilistic confidence.

\begin{figure}[t]
    \centering
    \includegraphics[width=\textwidth]{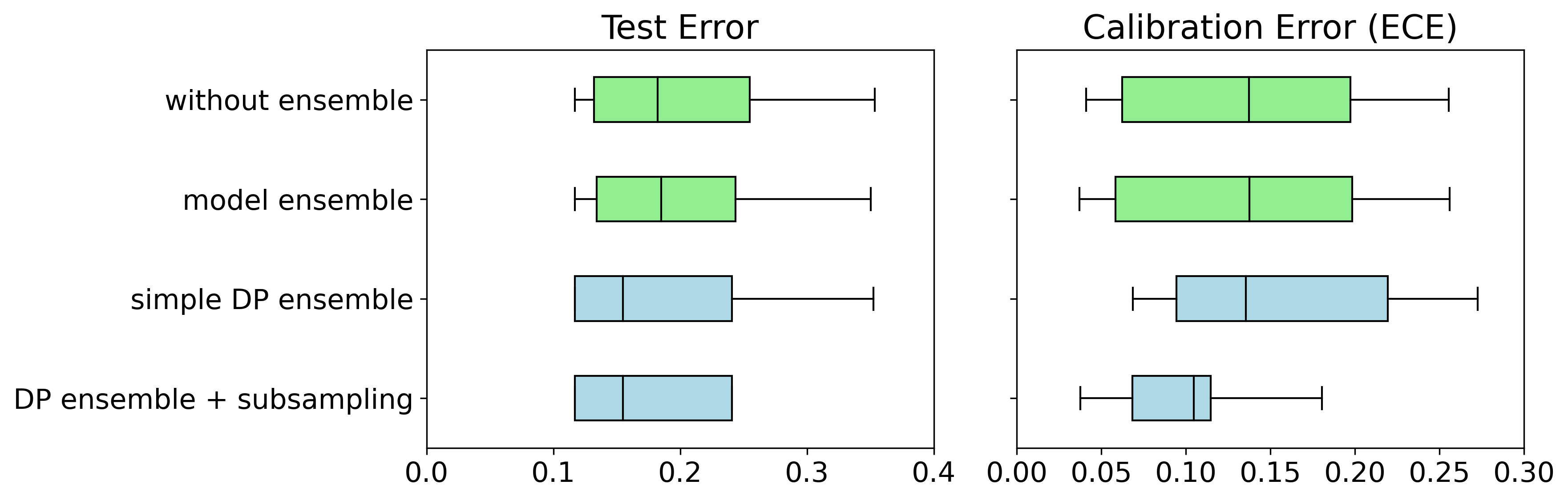}
    \caption{Comparison of downstream performance of various ensemble strategies on GAN-based DP mechanisms. 
    }
    \label{fig:gan}
    \vspace{2em}
\end{figure}

\vspace{-0.5em}
\paragraph{Results.}
For every dataset, DP mechanism, and downstream model class, we calculate the test error and calibration error when training on private synthetic data and testing on real data. The results are visualized using a box-and-whisker plot for each ensemble strategy proposed in Section~\ref{sec::ensemb_strategy}. We present these results for marginal/workload-based mechanisms in Figure~\ref{fig:marginal} and for GAN-based mechanisms in Figure~\ref{fig:gan}. Our key observations are: 

\begin{figure}[t]
    \centering
    \includegraphics[width=\textwidth]{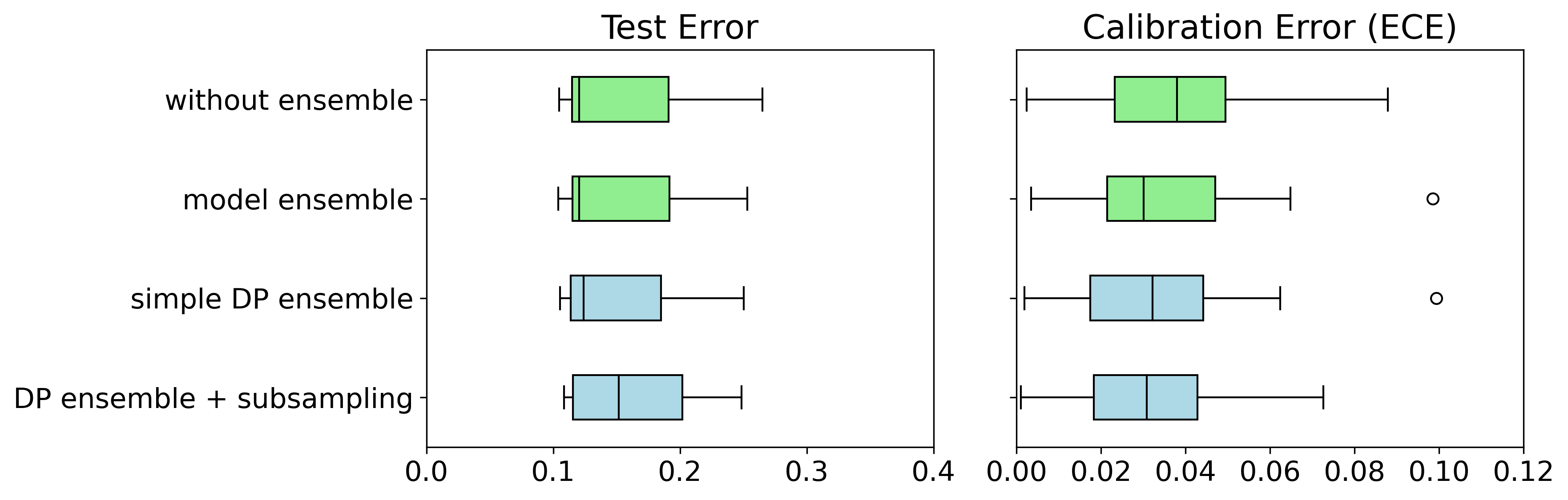}
    \caption{Comparison of downstream performance of various ensemble strategies on marginal-based and workload-based DP mechanisms.}
    \label{fig:marginal}
\end{figure}

\vspace{-0.5em}
\begin{itemize}[leftmargin=*]
    \item \textit{Model ensemble} exhibits little improvement in calibration or test errors compared with \textit{without ensemble}. In other words, when using a single synthetic dataset, there appears to be no significant advantage in training an ensemble of models over training a single model. This observation indicates that the distribution shift between synthetic and real data plays a more pivotal role in influencing the performance of downstream models.

    \item When synthetic data is generated using marginal-based or workload-based mechanisms, none of the ensemble strategies from Section~\ref{sec::ensemb_strategy} outperform the \textit{without ensemble} approach. 
     In the \textit{DP ensemble} method, while we achieve greater diversity in synthetic data by applying the DP mechanism multiple times, each invocation receives a smaller privacy budget, leading to a more pronounced distribution shift. 
     For marginal-based or workload-based mechanisms, these two factors cancel out and lead to no improvements in the downstream models.

    \item When synthetic data is generated using GAN-based mechanisms, \textit{DP ensemble + subsampling} outperforms all other strategies in terms of calibration and test error. 
    It indicates that subsampling is particularly beneficial in inducing greater diversity for GAN-based mechanisms.

\end{itemize}
\vspace{0.25em}

\section{Conclusion and Future Work}

DP synthetic data plays a pivotal role in enabling companies to share data for analytical or ML tasks without risking privacy breaches. However, synthetic data can be noisier than the original data and follow a different probability distribution due to the noise introduced to ensure DP. Consequently, ML models trained on synthetic data may suffer from a performance reduction when applied to real data. In this paper, we introduce several ensemble strategies for training downstream models. Through comprehensive numerical experiments, we found that (i) for synthetic data generated from marginal-based and workload-based mechanisms, ensemble strategies do not offer notable improvements compared to models trained \emph{without ensemble}; (ii) for synthetic data generated from GAN-based methods, \emph{DP ensemble + subsampling} outperforms all other strategies in terms of calibration and test error when training downstream models.

We believe there is a crucial need for future research to understand how ensemble strategies impact the performance of downstream models. For example, it would be interesting to provide an impossibility result that theoretically explains why ensemble strategies do not enhance downstream model performance for marginal-based  and workload-based mechanisms. 
Additionally, one can potentially further improve our ensemble strategies by reserving a hold-out set from real data and using it to select synthetic data generated by different invocations of the privacy mechanism.

Another promising direction for future research is to develop a more robust learning paradigm that trains ML models with synthetic data while taking into account potential distribution shifts. Existing studies \citep[see, e.g.,][]{tao2021benchmarking,mckenna2021winning,aydore2021differentially,nikolov2013geometry} have found that given sufficient data, marginal-based and workload-based mechanisms are effective in preserving statistical properties of the real data distribution, particularly those characterized by workload queries. It would be interesting to leverage this property and apply distributional robust optimization (DRO) to train ML models that can maintain their performance upon deployment on real data.

\bibliographystyle{IEEEtran}
\bibliography{reference}

\newpage
\appendix
\onecolumn

\appendix

\section{Details on the Experimental Results}
\label{append::exp}

\subsection{Benchmark Datasets}
\label{append:data}

\begin{table}[h]
    \centering
    \begin{tabular}{l|c c c}
    \hline
         Name &  Size & Categorial & Numerical\\
         \hline
         Adult & 48842 & 9 & 6\\
         Bank Marketing & 45211 & 9 & 7\\
         Online Shoppers & 12330 & 7 & 10\\
         \hline
    \end{tabular}
    \caption{Specification of the datasets used in our study.}
\end{table}

\subsection{Details on the Classification Models}
We used the \texttt{scikit-learn} library \cite{scikit-learn} to implement all models used in our experiments.
All parameters used the library default values (as of version 1.3.0) unless otherwise noted.
\begin{enumerate}[leftmargin=*]
    \item For logistic regression, we used the \texttt{LogisticRegression} class.
    \item For random forest, we used the \texttt{RandomForestClassifier} class. For the random forest's maximum depth, we used both the default value and \texttt{max\_depth = 5}.
    \item For gradient boosted decision tree, we used the \texttt{HistGradientBoostingClassifier} class.
    \item Because directly computing the kernel of a SVM is computationally prohibitive, we approximate the RBF kernel with random Fourier features. This is defined by \texttt{make\_pipeline(RBFSampler(gamma='scale'), SGDClassifier(loss='hinge'))}. Then, we compute the Platt scaling with the \texttt{CalibratedClassifierCV} class.
\end{enumerate}

\subsection{Additional Experimental Results}
\label{append::result}
Recall that $\epsilon$ is the total privacy budget and $k$ is the size of the ensemble.
We also ran the experiments for combination of $\epsilon \in \{3.0, 5.0\}$ and $k \in \{3, 5\}$.
The results of these additional experiments are largely similar to those of Figures~\ref{fig:marginal} and~\ref{fig:gan}, for which we used $\epsilon = 3, k = 3$.

\begin{figure}[!htb]
    \centering
    \includegraphics[width=\textwidth]{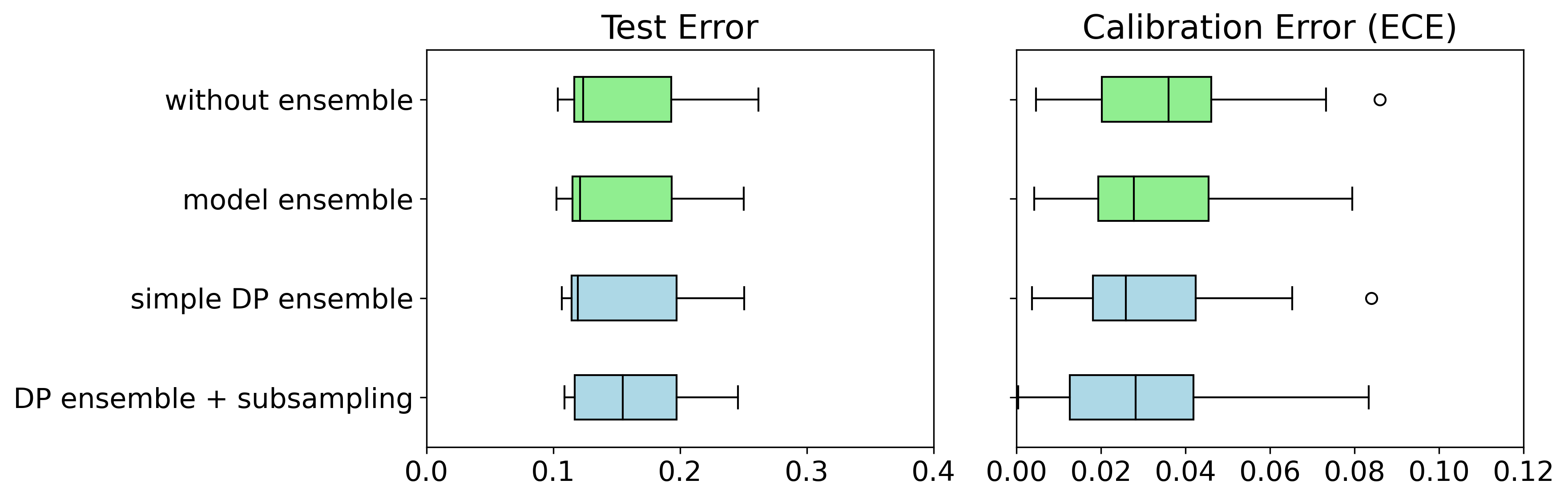}
    \caption{Comparison of downstream performance of various ensemble strategies on marginal-based and workload-based DP mechanisms for $\epsilon = 3.0, k = 5$.}
\end{figure}

\begin{figure}[!htb]
    \centering
    \includegraphics[width=\textwidth]{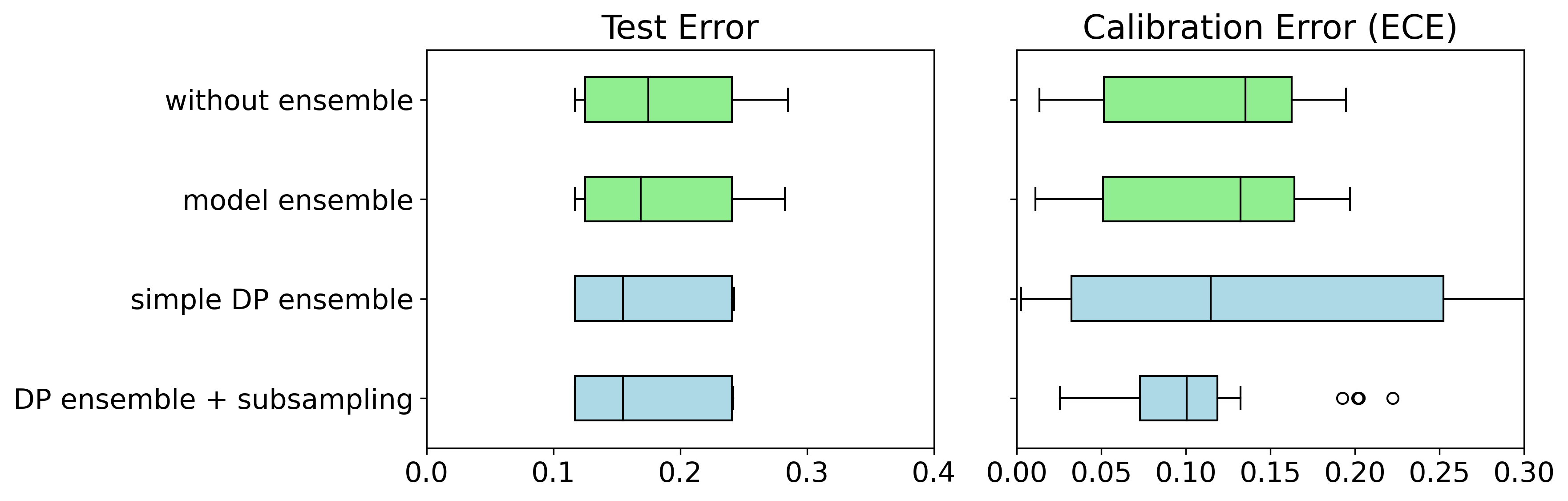}
    \caption{Comparison of downstream performance of various ensemble strategies on GAN-based DP  mechanisms for $\epsilon = 3.0, k = 5$.}
\end{figure}

\begin{figure}[!htb]
    \centering
    \includegraphics[width=\textwidth]{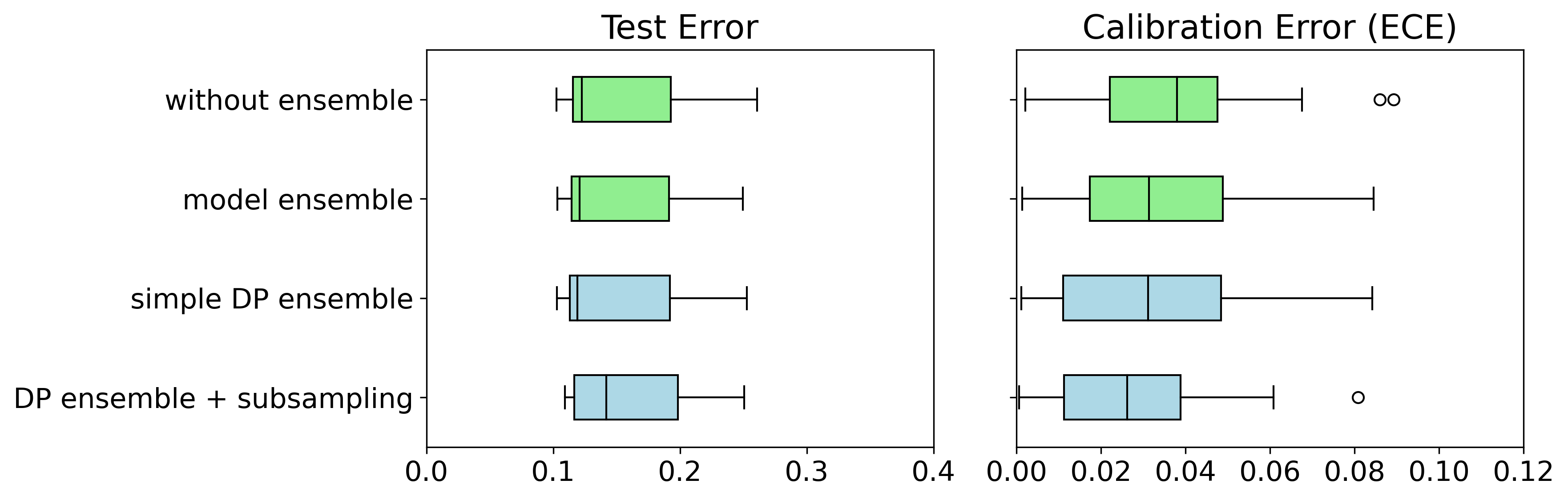}
    \caption{Comparison of downstream performance of various ensemble strategies on marginal-based and workload-based DP mechanisms for $\epsilon = 5.0, k = 3$.}
\end{figure}

\begin{figure}[!htb]
    \centering
    \includegraphics[width=\textwidth]{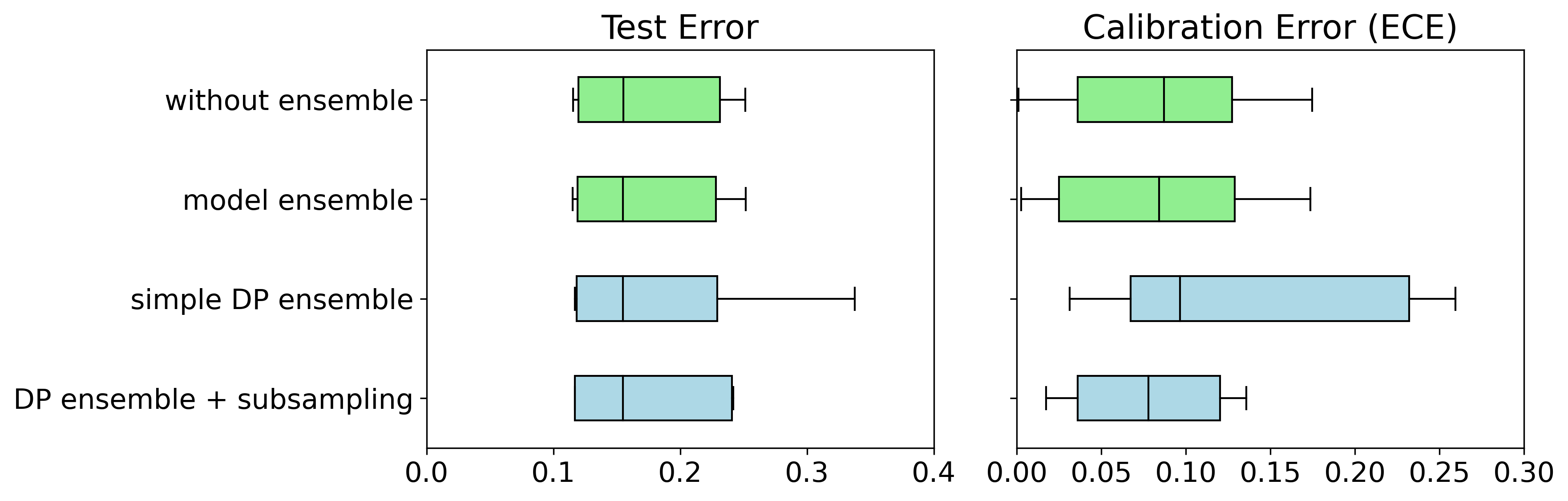}
    \caption{Comparison of downstream performance of various ensemble strategies on GAN-based DP  mechanisms for $\epsilon = 5.0, k = 3$.}
\end{figure}

\begin{figure}[!htb]
    \centering
    \includegraphics[width=\textwidth]{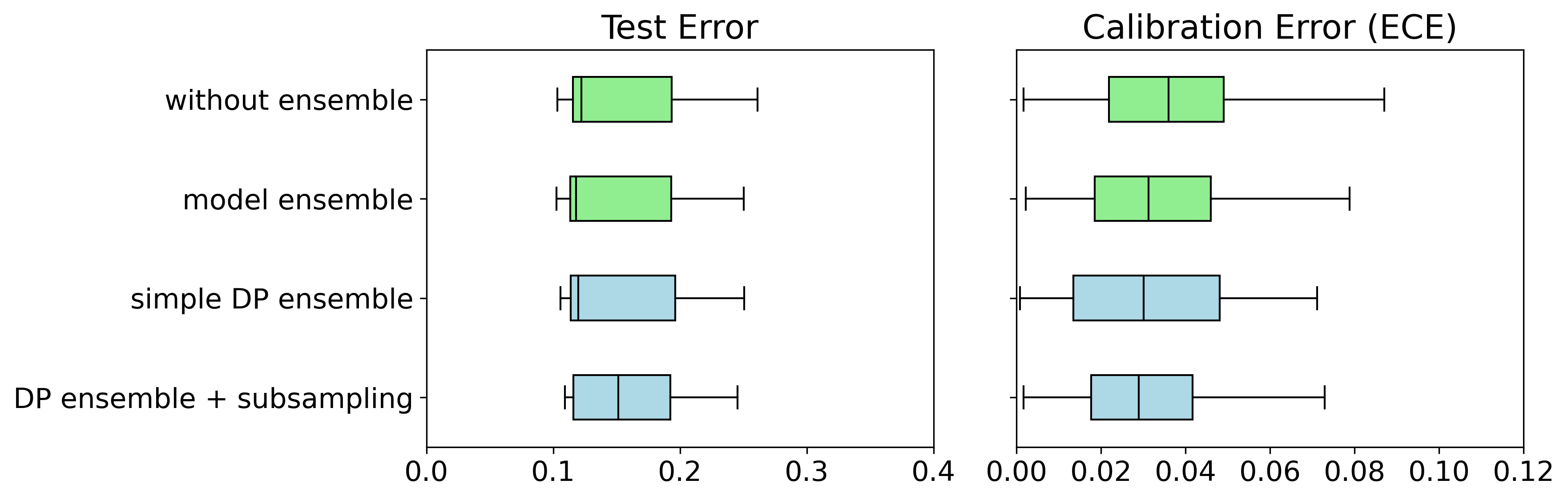}
    \caption{Comparison of downstream performance of various ensemble strategies on marginal-based and workload-based DP mechanisms for $\epsilon = 5.0, k = 5$.}
\end{figure}

\begin{figure}[!htb]
    \centering
    \includegraphics[width=\textwidth]{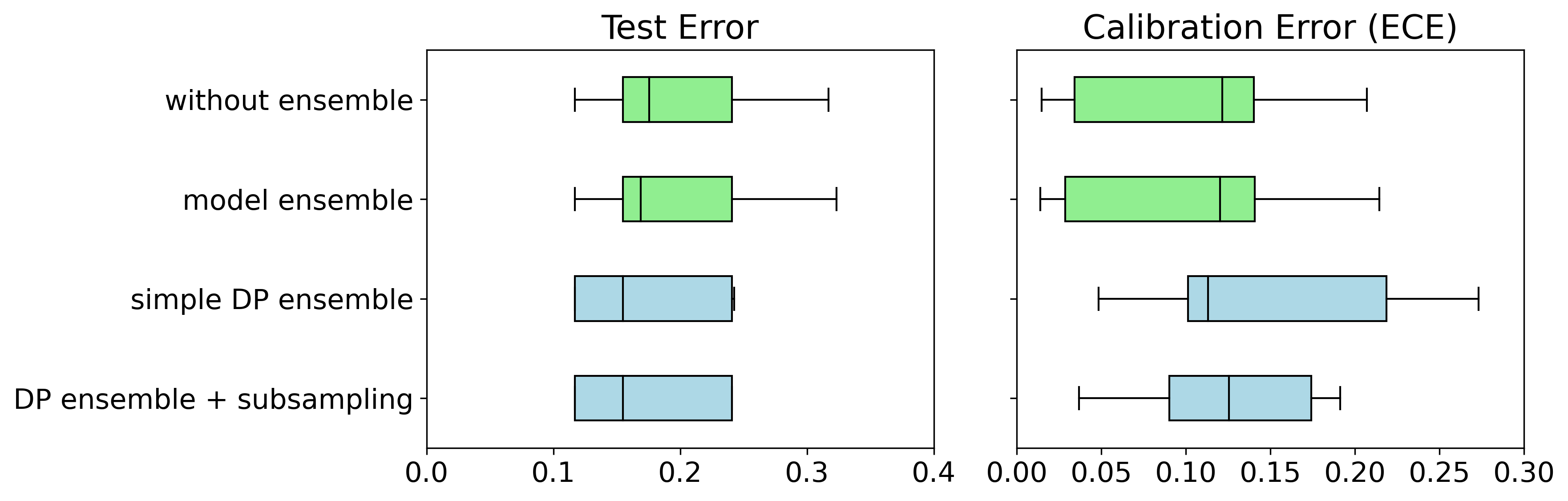}
    \caption{Comparison of downstream performance of various ensemble strategies on GAN-based DP  mechanisms for $\epsilon = 5.0, k = 5$.}
\end{figure}

\end{document}